\documentclass[twocolumn]{emulateapj}

\newcommand{\Ha}{H$\alpha$}

\newcommand{\arp}{Arp\,102B}

\newcommand{\OIww}{\hbox{[O\,{\sc i}]$\lambda\lambda $6300,6364}}
\newcommand{\Haw}{\hbox{H\,$\alpha\lambda $6563}}

\newcommand{\calacs}{{\sc Calacs}}

\def\flux{erg$^{-1}$ cm$^{-2}$ \AA$^{-1}$}
\def\kms{$\mbox{km s}^{-1}$}
\def\kmsmpc{$\mbox{km s}^{-1}\mbox{ Mpc}^{-1}$}
\def\perpix{$\mbox{pixel}^{-1}$}

\def\deg{^\circ}

\slugcomment{Accepted for publications in The Astrophysical Journal}

\shorttitle{An H$\alpha$ nuclear spiral structure in Arp\,102B}
\shortauthors{Kambiz Fathi et al.}

\begin{document}
\title{An H$\alpha$ nuclear spiral structure in the E0 active galaxy Arp\,102B}
\author{Kambiz Fathi\altaffilmark{1,2,3*}}
\author{David J. Axon\altaffilmark{3,4}}
\author{Thaisa Storchi-Bergmann\altaffilmark{5}}
\author{Preeti Kharb\altaffilmark{3}}
\author{Andrew Robinson\altaffilmark{3}}
\author{Alessandro Marconi\altaffilmark{6}}
\author{Witold Maciejewski\altaffilmark{7}}
\author{Alessandro Capetti\altaffilmark{8}}
\altaffiltext{1}{Stockholm Observatory, Department of Astronomy, Stockholm University, AlbaNova Centra, SE-106 91 Stockholm, Sweden. * Major part of this work was completed at RIT.}
\altaffiltext{2}{Instituto de Astrof\'\i sica de Canarias, 38200 La Laguna, Spain. 
\\Email: fathi@iac.es}
\altaffiltext{3}{Department of Physics, Rochester Institute of Technology, 
85 Lomb Memorial Dr., Rochester, NY 14623, USA.
\\Email: djasps@rit.edu ; axrsps@rit.edu, pxksps@cis.rit.edu}
\altaffiltext{4}{Department of Physics and Astronomy, University of Sussex, UK}
\altaffiltext{5}{Instituto de F\`\i sica, UFRGS, Av. Bento Goncalves 9500,
91501-970 Porto Alegre RS, Brazil.
\\Email: thaisa@ufrgs.br ; rogemar@ufrgs.br}
\altaffiltext{6}{INAF - Osservatorio Astrofisico di Arcetri,
Largo Fermi 5, I-50125 Firenze, Italy.
\\Email: marconi@arcetri.astro.it}
\altaffiltext{7}{Astrophysics Research Institute, Liverpool John Moores University, Twelve
Quays House, Egerton Wharf, Birkenhead, CH41 1LD, UK.
\\Email: wxm@astro.livjm.ac.uk}
\altaffiltext{8}{INAF - Osservatorio Astronomico di Torino,
Strada Osservatorio 20, 10025 Pino Torinese, Italy.
\\Email: capetti@to.astro.it}

\begin{abstract}
We report the discovery of a two-armed mini-spiral structure within the inner kiloparsec of the E0 LINER/Seyfert 1 galaxy \arp. The arms are observed in \Ha\  emission and located East and West of the nucleus, extending up to $\approx$1\,kpc from it. We use narrow-band imaging from the Hubble Space Telescope Advanced Camera for Surveys, in combination with archival Very Large Array radio images at 3.6 and 6 cm to investigate the origin of the nuclear spiral. {From the \Ha\ luminosity of the spiral, we obtain an ionized gas mass of the order of $10^6$ solar masses.} One possibility is that the nuclear spiral represents a gas inflow triggered by  a recent accretion event which has replenished the accretion disk,  giving rise to the double-peaked emission-line profiles characteristic of \arp.  However, the radio images show a one-sided curved jet which correlates with the eastern spiral arm observed in the \Ha\  image. A published milliarcsecond radio image also shows one-sided structure at position angle $\approx$\,40$\deg$, approximately aligned with the inner part of the eastern spiral arm. The absence of a radio counter-part to the western spiral arm is tentatively interpreted as indicating that the jet is relativistic, with an estimated speed of 0.45$c$. Estimates of the jet kinetic energy and the ionizing luminosity of the active nucleus indicate that both are capable of ionizing the gas along the spiral arms. {We conclude that, although the gas in the nuclear region may have originated in an accretion event, the mini-spiral is most likely the result of a jet-cloud interaction rather than an inflowing stream.}
\end{abstract}


\section{Introduction}
Over the past few years, observations of the central regions of {active} galaxies have shown that nuclear structures (within the inner kpc), such as bars, rings and spirals seem to be ubiquitous (e.g., Pogge \& Martini 2002; Fathi 2004; Jogee et al. 2005). The most common nuclear structures are dusty spirals which are estimated to reside in more than half of all active galaxies. The enhanced frequency of nuclear spirals in active galaxies as compared to non-active galaxies (e.g., Martini et al. 2003), and in particular in early-type hosts (Sim\~oes Lopes et al. 2007) supports the hypothesis that nuclear spirals are a mechanism for fueling the nuclear supermassive black hole (hereafter SMBH). {This has now been confirmed by a few studies such as in the LINER/Seyfert 1 galaxy NGC\,1097 (Fathi et al. 2006; Davies et al. 2009; van de Ven \& Fathi 2010), in the LINER galaxy NGC\,6951 (Storchi-Bergmann et al. 2007), NGC\,4051 (Riffel et al. 2008), NGC\,1068 (Mueller Sanchez et al. 2009) and M\,81 (Schnorr M\"uller et al. 2011)}, using two-dimensional spectroscopy together with high resolution imaging to map the nuclear spiral arms. These authors have measured the gas emission-line kinematics in the spiral arms, confirming gas streaming from scales of hundred of parsecs towards the nucleus.

Morphologically, nuclear spirals are divided into two main categories, chaotic spirals, and symmetric or grand design spirals, for which different formation mechanisms have been proposed. Chaotic spiral arms could be formed by acoustic instability (Elmegreen et al. 1998; Montenegro et al. 1999),  whereas grand design spirals could form by density waves (Engelmaier \& Shlosman 2000; Saijo, Baumgarte \& Shapiro 2003) or by hydrodynamic shocks induced by non-axisymmetric potentials (Maciejewski et al. 2002).  Martini \& Pogge (1999) have shown that nuclear spirals are not  self-gravitating, and that they are likely to be shocks in the  nuclear gas disks (e.g., Yuan \& Yen 2004). Maciejewski (2004a,b) has demonstrated that, if a central SMBH is present, spiral shocks can extend all the way to its vicinity and generate gas inflow compatible to the accretion rates observed in local active galactic nuclei (AGN).

The nuclear spirals are distinct from the spiral arms in the disks of galaxies as they seem to have lower density contrasts and no associated star formation. Possible ionization mechanisms for the emission-line gas observed (e.g. in NGC\,1097; Storchi-Bergmann et al. 1996; Fathi et al. 2006, or NGC\,2974; Emsellem et al. 2003) are the spiral  shocks, normally with velocities of the order of $\approx 100$ \kms\ and photoionization by an active nucleus. 

Here we report the discovery of a nuclear spiral in the inner kiloparsec of the LINER/Seyfert 1 galaxy \arp. Although mini spirals have been observed in other galaxies, as pointed out above, the case of \arp\ is particularly interesting  because the host galaxy is a ``featureless" E0 galaxy. Another peculiar  characteristic of \arp\ is the presence of very broad ($\approx$15,000~\kms) double-peaked emission-lines in its nuclear spectrum, similar to what is observed in NGC\,1097 (Storchi-Bergmann et al. 1993, 2003; Fathi et al. 2006) and also in some other LINERs -- e.g. M\,81 (Bower et al. 1996), NGC\,4203 (Shields et al. 2000), NGC\,4450 (Ho et al. 2000) and NGC\,4579 (Barth et al. 2001), as well as in radio galaxies (e.g. Eracleous \& Halpern 1994, Lewis, Eracleous \& Storchi-Bergmann 2010). In fact, \arp\ is considered to be the prototypical ``double-peaker" (Chen, Halpern \& Filippenko 1989; Newman et al. 1997; Gezari, Halpern \& Eracleous 2007; Flohic \& Eracleous 2008).

We speculate that the appearance of double-peaked lines is related to a recent accretion event. This seems to be the case of NGC\,1097, M\,81 and of some of the other LINERs above, which had not shown double-peaked lines in previous observations. It thus may be that the nuclear spiral in \arp\  is also a tracer of this accretion event. As noted above, nuclear spirals are also present in NGC\,1097, M\,81 and other LINERs with double-peaked lines.

In this work we present high resolution Hubble Space Telescope images acquired with the Advanced Camera for Surveys ({\it HST}-ACS) in both the narrow-band \Ha\ and the continuum. These data are described in sections \ref{sec:hstdata} and \ref{sec:hstresults}. In section \ref{sec:vladata} we present unpublished archival Very Large Array (VLA) radio images;  in section \ref{sec:discussion} we compare the {\it HST} and VLA data discussing the relationship between the radio structure and the spiral arms of the galaxy and in section \ref{sec:conclusions} we present our conclusions.

\begin{figure*}
\centering{
\includegraphics[width=.99\textwidth]{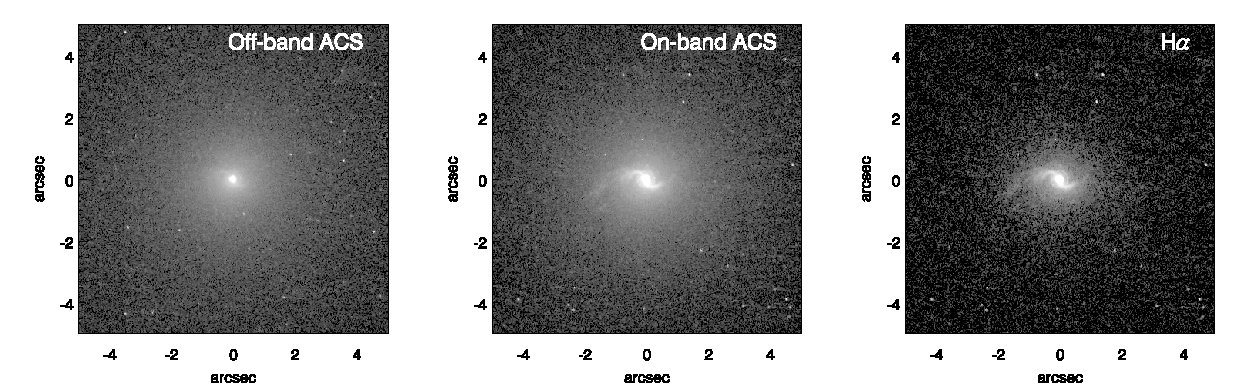} 
\caption{From left to right, our {\em ``off-band''} {\it HST}-ACS continuum image, 
{\em ``on-band''} {\it HST}-ACS ramp-filter image, and continuum 
subtracted \Ha\ emission-line image. 
In all panels north is up and east is to the left.}}
\label{fig:HSTfull}
\end{figure*}

\section{HST Observations}
\label{sec:hstdata}
{\it HST} data (GO~9782; PI: David J. Axon) were obtained as part of a larger program to study circum-nuclear gas kinematics in objects with double-peaked broad lines. {\it HST}-ACS images of \arp\ were obtained using the high resolution camera (0\farcs025 \perpix\ and a field of view of $26\arcsec\times29\arcsec$), through the FR656N filter (for the continuum) and the 6722\AA\ linear ramp filter centered on the \Haw\ emission-line at the redshift ($z=0.0242$) of \arp. The ACS images were reduced and flux calibrated using the standard \calacs\ pipeline (Pavlovsky et al. 2005). After reduction, the scaled continuum image was subtracted from the linear ramp filter image in order to isolate the pure \Ha\ emission (see Fig.~\ref{fig:HSTfull}). {In order to remove the point spread function (PSF) distortion and unveil the underlying structure in the ACS image, we deconvolved the image using 50--150 Richardson-Lucy iterations (Richardson 1972; Lucy 1974) in conjunction with a PSF model constructed using Tiny Tim (Krist \& Hook 1999). Different numbers of iterations produced similar results, and in all cases the final resolution of the image was comparable to that of the original ACS image (i.e. 0$\farcs$05), all showing a smooth underlying flux distribution with very prominent nuclear spiral arms in H$\alpha$ emission.}

\begin{figure}
\centering{
\includegraphics[width=.49\textwidth]{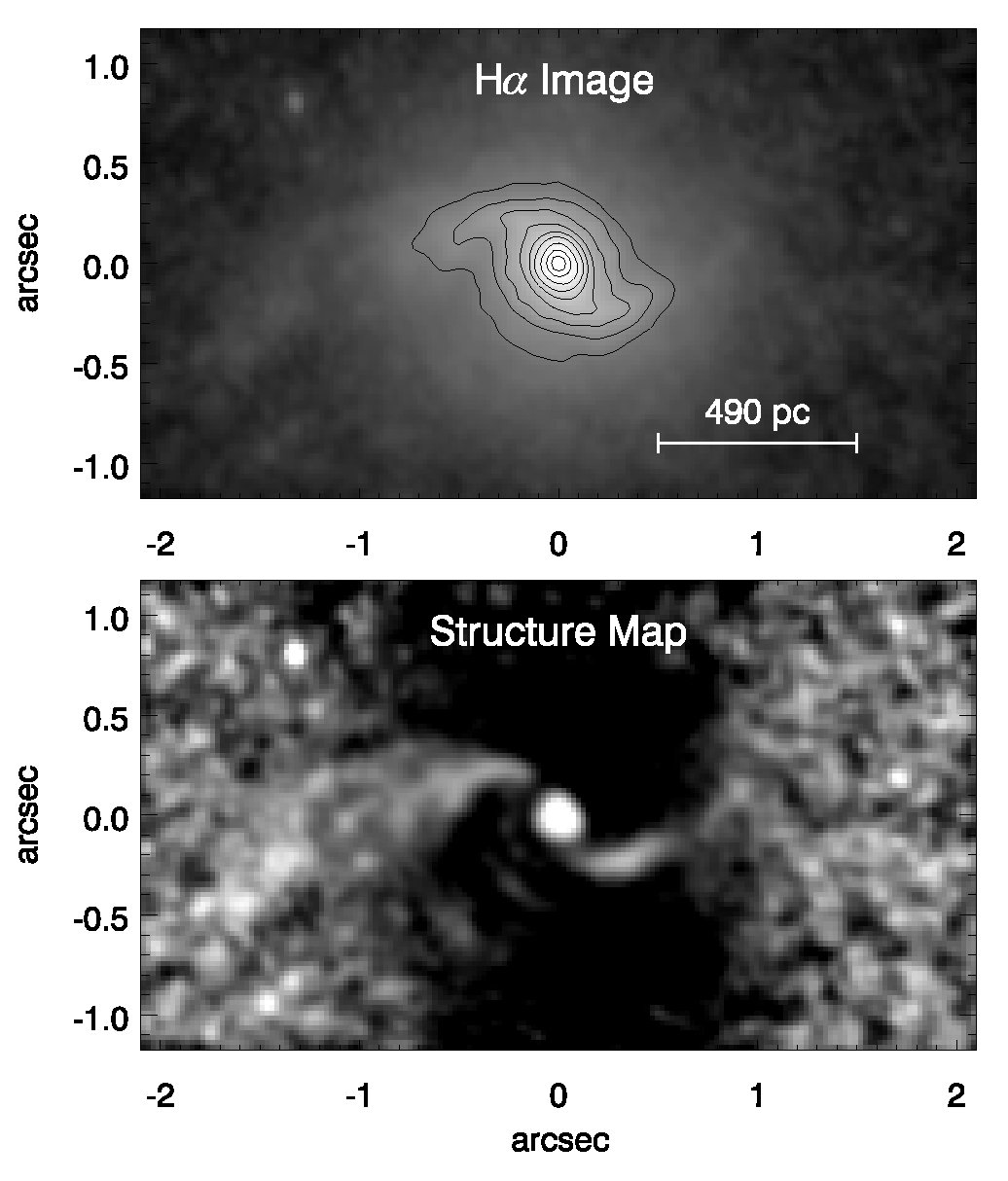} 
\caption{Zooming into the central few hundreds of parsecs of \arp. 
Top: \Haw\ image; Bottom: the structure map obtained from this image. 
Both panels have the same orientation of Fig.~\ref{fig:HSTfull}.}}
\label{fig:HSTzoom}
\end{figure}

\section{HST Results}
\label{sec:hstresults}

In the central and right panels of Fig.~\ref{fig:HSTfull}, the \Ha\ ramp filter and continuum subtracted \Ha\ images show two small spiral arms, one to the east and the other to the west of the nucleus. The arm to the west extends to $\approx$\,1\arcsec\, while the one to the east extends to $\approx$\,2\arcsec\ from the nucleus; in the outer parts, the east arm shows a broader and more ``fragmented'' structure. In Fig.~\ref{fig:HSTzoom}, we show the central $4\arcsec\times2\arcsec$ section of the continuum subtracted \Haw\ image together with a structure map of this image, built in order  to increase the contrast and better reveal the spiral morphology.

The structure map technique was proposed by Pogge \& Martini (2002), and is based on the Richardson-Lucy image restoration algorithm (Richardson 1972; Lucy 1974) applied to HST imaging (e.g., Snyder, Hammoud, \& White 1993). The image generated is called structure map S and is given by:

\begin{equation}
	S = \left[ \frac{I}{I \otimes P} \right] P^t
\end{equation}

\noindent Where I is the original image, P is the model PSF, constructed with the TinyTim software (Krist \& Hook 1999), P$^t$ is the transpose of the model PSF and $\otimes$ is the convolution operator. This technique removes the large scale smooth distribution, highlighting both dust (dark regions) and emission structures (bright regions) with high fidelity. In the structure map, the spiral arms can be traced down to the central $0\farcs1$\, (corresponding to $\approx$\,50 parsecs (pc) at the galaxy\footnote{Using the systemic velocity of 7245\,\kms (de Vaucouleurs et al. 1991) and  $H_0 = 71$ \kmsmpc\ (Bennett et al. 2003).}).

\begin{figure*}
\centering{
\includegraphics[width=.99\textwidth]{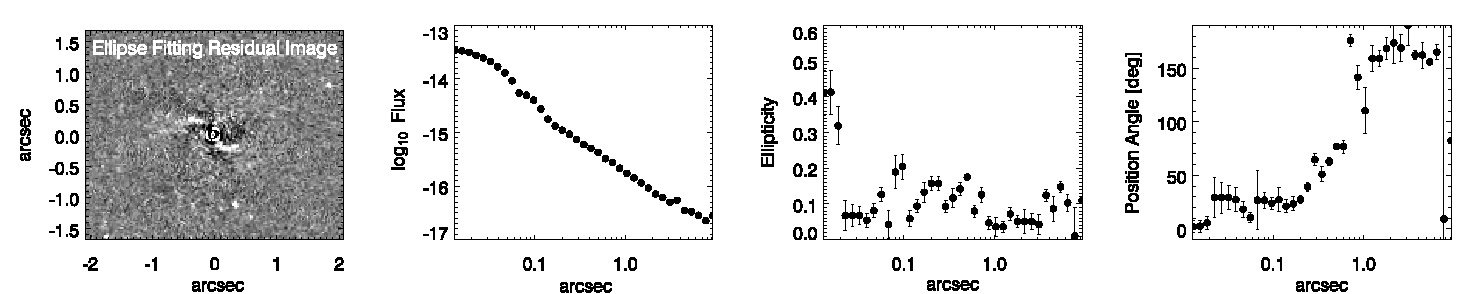} 
\caption{Ellipse fitting results for the continuum image of \arp. In the left panel we show the residual image obtained from the subtraction of a smooth elliptical model from the continuum image, where the flux is given  in \flux\ arcsec$^{-2}$. In the panels to the right we show the radial variation of the parameters of the ellipses.}}
\label{fig:HSTelfit}
\end{figure*}

In order to investigate the morphology of the continuum image, we fitted an elliptical model (concentric ellipses) to the inner 8\arcsec, and found a mean ellipticity of $0.098 \pm 0.059$. The results of this fit are shown in Fig.~\ref{fig:HSTelfit}. The position of the nucleus --  considered to be the center of the elliptical isophotes --  did not vary by more than 0.5 pixel. Within the fitted range, the position angle of the ellipses varies by almost $160\deg$, but the ellipticity is small and shows no systematic variations, suggesting that the position angle variation is not significant. The residual between the continuum image and the elliptical model (left panel of Fig.~\ref{fig:HSTelfit}) shows that the continuum also displays the spiral arms that we have found in the \Ha\ image. This suggests that the continuum image may not be displaying the pure stellar component, including some contribution from nebular emission lines, {in particular the \OIww\ doublet. An estimate of the contribution of the [OI] emission to the continuum image can be obtained from the [OI]$\lambda$6300 line flux from the ground-based optical spectrum of \arp\ included in the work by Halpern et al. (1996). This spectrum was obtained through an aperture of width  2\arcsec\ and length  4\arcsec, oriented at PA=80$\degr$, which approximately matches the size and orientation of the H$\alpha$ mini-spiral, and have emission-line  fluxes of 1.9$\times$10$^{-14}$erg\,cm$^{-2}$\,s$^{-1}$ and 4.3 $\times$10$^{-14}$erg\,cm$^{-2}$\,s$^{-1}$ for [OI]$\lambda$6300 and the narrow component of H$\alpha$, respectively, thus corresponding to  [OI]/H$\alpha$=0.45. Considering that our narrow-band H$\alpha$ image includes also contribution from the [NII]$\lambda\lambda$6548,84 doublet, an estimate for the ratio  between the residual image (continuum$-$elliptical model), under the assumption that this residual is due to [OI] emission, and the continuum subtracted H$\alpha$ image should be of the order of 0.2. This is indeed the approximate value we obtain for the ratio between the two images in the arms, supporting that the residual arms in the continuum are indeed due to contamination by the [OI]$\lambda$6300 emission line.}

We have integrated the flux in the continuum-subtracted \Ha\ image obtaining  
2.31\,$\times\,10^{-13}$\,erg\,cm$^{-2}$s$^{-1}$, but this includes the 
double-peaked and broad components of the \Ha\ line, which contribute with 
most of the flux in the inner pixels. Performing aperture photometry within a 
radius of 5 pixels around the nucleus, we estimate that the contribution of 
the nuclear (broad and double-peaked) components to the total \Ha\ flux is at 
least 50\%. Considering, in addition, that the filter includes also the 
[NII]$\lambda\lambda$6548,84\,\AA~emission lines (which together have 
similar fluxes to that of narrow \Ha), we  estimate that the total flux in the 
narrow \Ha\ component is 5.8$\pm1.7\times10^{-14}$\,erg\,cm$^{-2}$s$^{-1}$. 
Within the uncertainties, this value is consistent with the one reported by 
Halpern et al. (1996), of 4.3\,$\times\,10^{-14}$\,erg\,cm$^{-2}$s$^{-1}$.
As we do not have a strong constraint from our \Ha\ image for the 
broad and double-peaked \Ha\ components contribution, we preferred to use the 
narrow-line flux of Halpern et al. (1996) to calculate the  \Ha\ luminosity of 
the mini-spiral (see Section 5.3).

\section{Radio Data}
\label{sec:vladata}

\subsection{VLA Observations}
We obtained archival Very Large Array A and B-array configuration data at 8\,GHz ($\lambda$3.6\,cm) and B-array configuration data at 5, 8 and 15\,GHz ($\lambda$6, 3.6 and 2\,cm, Experiment code AW230). The observations were carried out on February 28th, 1993. 3C~286 was the primary flux density calibrator, while J1740+52 was the phase calibrator for the experiment. The data were processed with the Astronomical Image Processing System (AIPS) software using the standard imaging and self-calibration procedures. The final radio images were made after several iterations of phase and amplitude self-calibrations. Table~2 lists the beam-sizes and $rms$ noise in the final radio images.

As is evident from the beam-sizes, only the A-array 8\,GHz image had sufficient spatial resolution to enable a comparison with our {\it HST} images. This is illustrated in Figure 4, where the top panel shows the contours of the B-array image and the bottom panel of the A-array image overploted on the \Ha\ structure map. A lot of the discussion ahead focuses on the A-array image.

\begin{table*}
\caption{Summary of VLA Observations. Col.1: Array configuration of the VLA, Col.2: Frequency band,
Col.3: Central frequency in GHz, Col.4: Synthesized beam size,
Col.5: Peak core surface brightness in mJy\,beam$^{-1}$,
Col.6: Total radio flux density in mJy,
Col.7: $rms$ noise in image in $\mu$Jy\,beam$^{-1}$.}
\centering \tabcolsep=15pt
\begin{tabular}{ccccccc}
\hline \hline
Config & Band &Frequency & Beam & Peak & Total & $rms$\\
\  & \  & (GHz) & (arcsec$^2$) & (mJy\,bm$^{-1}$) & (mJy) & ($\mu$Jy\,bm$^{-1}$) \\ \hline
A & X &  8 & 0.23$\times$0.19   & 256   &270& 27   \\
B & C &  5 & 1.45$\times$1.11   & 116   &135& 26   \\
B & X &  8 & 0.75$\times$0.61   & 216   &227 & 75  \\
B & U & 15 & 0.42$\times$0.33   & 288 &297 & 25  \\ \hline
\end{tabular}
\end{table*}

\subsection{Results from the VLA study}
\label{radio}

The radio image (Figure 4) shows a one-sided radio jet emerging from the core at a PA of $\approx 90\deg$ towards the east and then bending towards the south. {The first image of the radio jet was presented by Puschell et al. (1986), although they were then not clear about the radio emission mechanism. Subsequently, Caccianiga et al. (2001) have clearly demonstrated that the parsec-scale radio emission in Arp\,102B is from an AGN jet, rather than a starburst, on the basis of its high brightness temperature ($T_{b}\sim10^{6}-10^{8}$\,K). Starbursts typically reveal $T_{b}<10^{5}$\,K (Condon et al. 1991).} We find that the VLA jet shows two radio knots surrounded by low surface brightness emission. The bending to the south occurs at the first radio knot $\sim0\farcs8$ (corresponding to $\sim$400\,pc at the galaxy) from the radio core. Further, the jet seems to flare out after the second radio knot at a distance of $\sim1\farcs3$ ($\sim$\,650\,pc). {By taking a transverse surface brightness slice and fitting a Gaussian to it (with task SLFIT in AIPS), we found that the jet width increased from $\sim0\farcs25$ at a distance of 
$0\farcs7$, to $\sim0\farcs75$ at a distance of $1\farcs3$ away from the core.} There is some indication of the beginning of a faint counterjet to the west. However, more sensitive observations are required to confirm this feature. The simplest interpretation for the asymmetry is that the eastern jet is directed partly towards us and is Doppler boosted.

Further evidence for relativistic boosting of the jet comes from milliarcsecond-scale observations with the European VLBI Network (EVN), which detected the radio core and a single additional radio component to the North-East of the nucleus at a PA of $\approx 40\degr$ (Caccianiga et al. 2001). Since our A-array image  shows the jet emerging at PA $\approx90\degr$, the jet seems to have changed its trajectory considerably. Using the EVN image, we have estimated the jet-to-counterjet surface brightness ratio, $R_J$, at an approximate projected distance of $0\farcs02$\,(9.6\,pc) from the radio core. Using the peak surface brightness of the second radio component (see Caccianiga et al. 2001) and assuming the $rms$ noise to be an upper limit to the counterjet surface brightness, we derive an $R_J\ge13$. Assuming the jet structural parameter, $p=3$ ($p=2-\alpha$ for a continuous jet, $\alpha$ being the jet spectral index, defined as $S_{\nu}\propto\nu^{\alpha}$; Urry \& Padovani, 1995), and following the relation, $R_J$=((1+$\beta$cos$\theta$)/(1$-\beta$cos$\theta$))$^p$, we obtain $\beta$cos$\theta\ge$0.4, where $\theta$ is the angle of the radio jet with respect to the line of sight. If we assume that the radio jet is launched perpendicular to the accretion disk, which in turn has its axis inclined at 33$\degr$ (Chen \& Halpern 1989), we obtain a jet velocity of $v\ge0.48c$ at a distance of $\sim$10\,pc from the core. The radio jet {in Arp\,102B} therefore seems to start out relativistically. 
We note that proper motion studies have indeed detected similar (high) speeds in the parsec-scale jets of some Seyferts (e.g., Middelberg et al., 2004). {Giroletti et al. 2005 infer a jet speed of $\sim0.75c$ in the S-shaped parsec-scale jet of the LINER galaxy, 
NGC\,4278. The asymmetry in the jet intensity from the famous Seyfert galaxy, NGC\,1068, has also been proposed to result from mild relativistic boosting in its jet (e.g., Axon et al., 1998).}

\begin{figure*}
\centering{
\includegraphics[width=.69\textwidth]{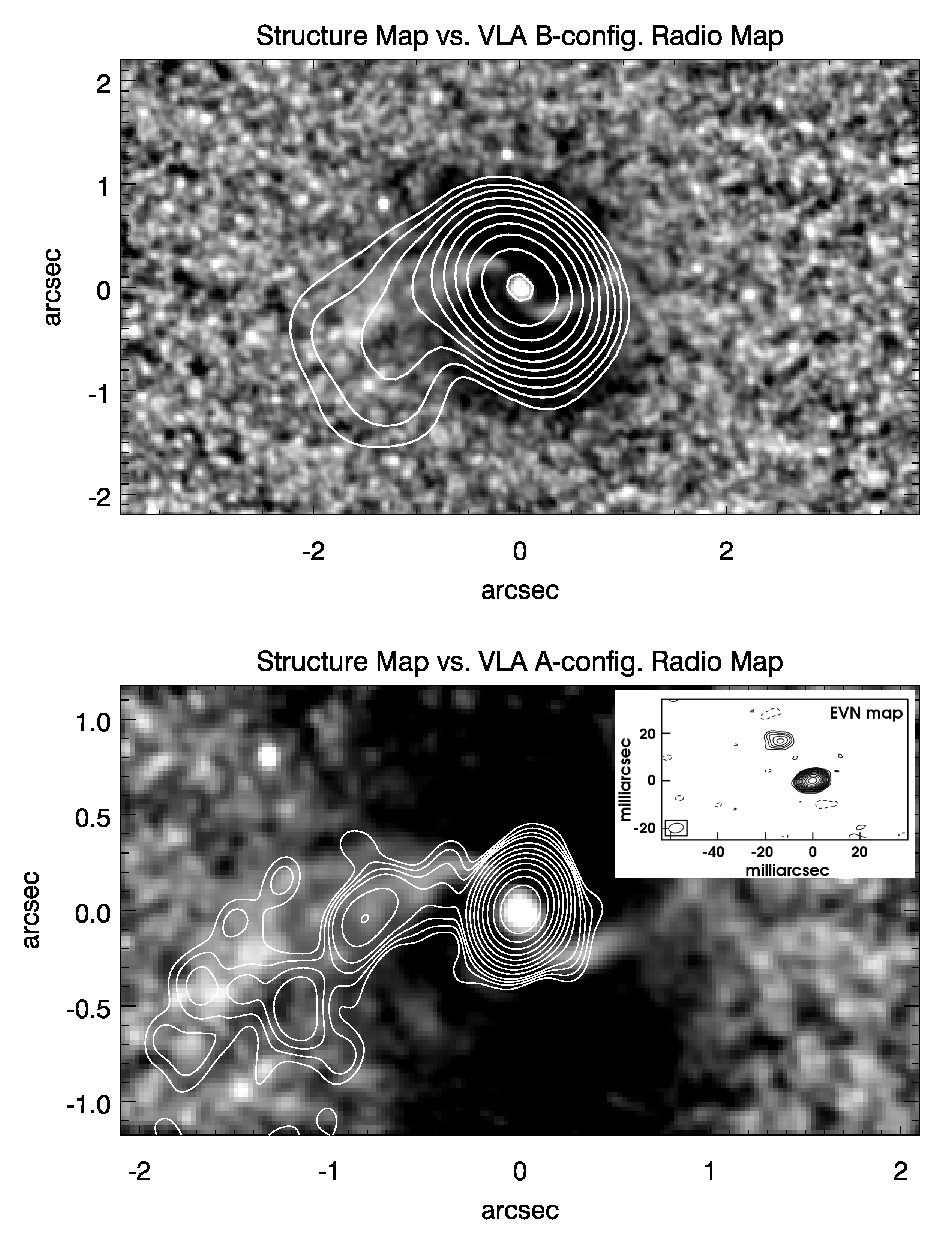} 
\caption{Our {\it HST}-ACS \Ha\ structure map with overploted {8\,GHz} contours of the  VLA B-configuration (top panel) and A-configuration (bottom panel) radio images. The radio contours cover a range of (0.08 $-$ 90\%) of the peak radio flux, {while the beam-sizes are $0\farcs75\times0\farcs61$ at PA = $43\degr$ in the top panel, and $0\farcs24\times0\farcs19$ at PA = $-18\degr$, in the bottom panel, respectively.} The lowest contour indicates a 4-$\sigma$ detection level. The inset image of the inner few tens of milliarcseconds is taken from Caccianiga et al. (2001),
and displays the radio core and the radio component at position angle $40\degr$.}}
\label{fig:HSTradio}
\end{figure*}

\begin{figure}
\begin{center}
\includegraphics[width=.49\textwidth]{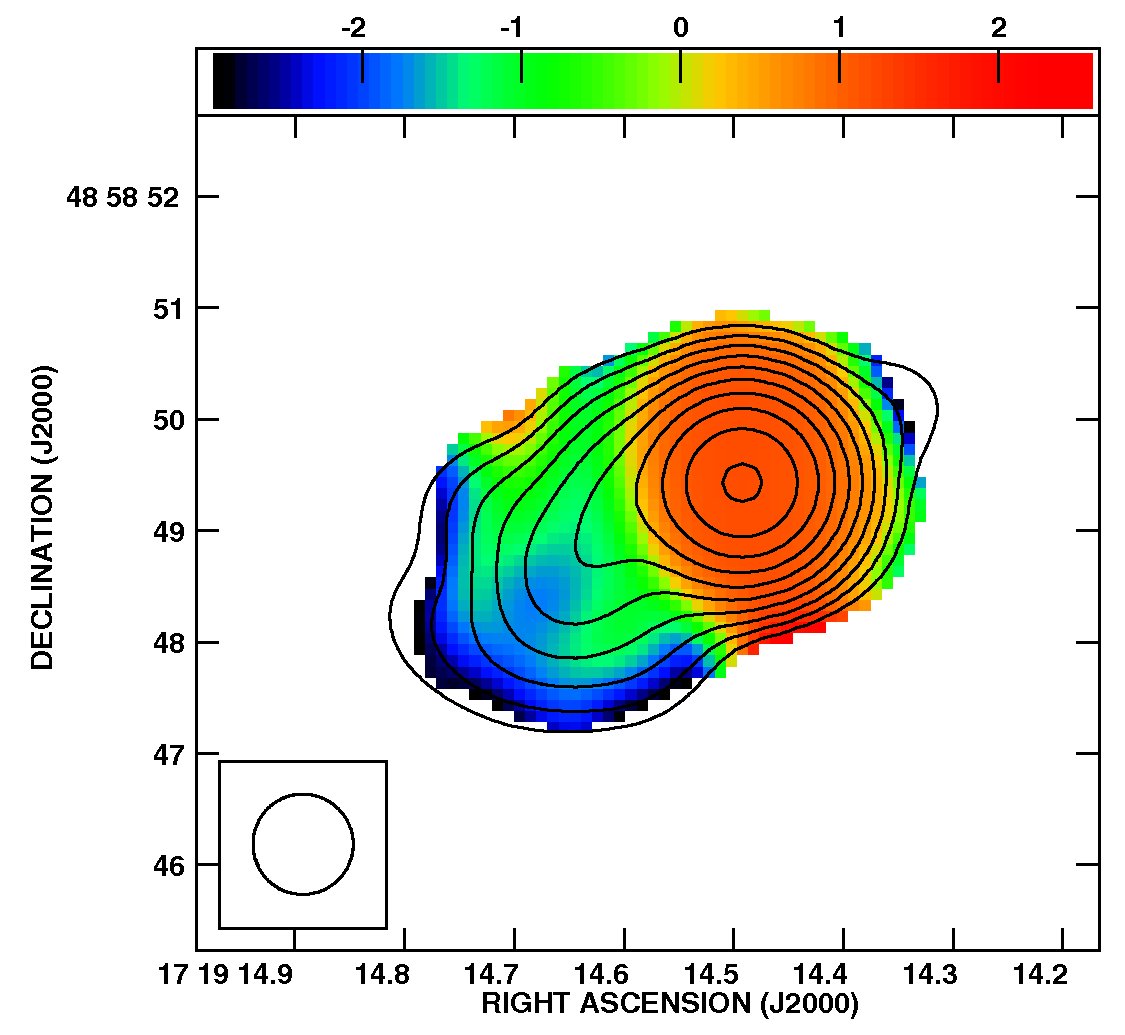} 
\caption{The 5$-$8\,GHz spectral index map of Arp\,102B made with
a $0\farcs9$ circular beam. The 5\,GHz total intensity contours are superimposed.}
\label{fig:spectral}
\end{center}
\end{figure}

We constructed a spectral index image using the B-array 5\,GHz and 8\,GHz radio images, after first convolving both with a circular beam of size $0\farcs9\times0\farcs9$ (Fig.~\ref{fig:spectral}). We found that while the radio core has an inverted spectral index {($\alpha=1.15\pm0.001$)}, and the inner jet has a steep spectrum {($\alpha=-1.04\pm0.01$ at a distance $<1\arcsec$ from the core)}, the spectrum steepens considerably in the outer part of the radio structure {($\alpha=-1.95\pm0.14$, at a distance of $\sim1\farcs5$ from the core)}. 
Although the northern edge of the jet shows a clear spectral flattening {($\alpha=-0.46\pm0.15$)} at a distance of $\sim1\farcs5$ from the radio core (see top panel of Figure 4), which could be indicative of a jet shock, the spatial resolution in the spectral index image is not sufficient to make a one-to-one comparison with features in the {\it HST} image. The 15\,GHz B-array image could resolve only the inner jet structure ($<1\arcsec$ from the core) and the 8$-$15\,GHz spectral index image did not reveal any more detail than the 5$-$8\,GHz spectral index image. The overall radio 
morphology and the 5$-$8\,GHz spectral index image could however be considered as being consistent with the picture of  a radio jet impacting on emission-line clouds in the nuclear ISM.

Under the assumption of ``equipartition'' of energy between relativistic particles and the magnetic field (Burbidge 1959), we estimated the minimum magnetic field strength and other parameters for a cylindrical jet geometry. The total radio luminosity was estimated assuming that the radio spectrum extends from 10\,MHz to 10\,GHz with a spectral index of $\alpha=-1$. Furthermore, it was assumed that the relativistic protons and electrons have equal energies, and the radio emitting plasma has a volume filling factor of unity. From the 8\,GHz image, we estimate the integrated jet flux density to be $\sim270$~mJy and the size of the radio jet to be $2\farcs5\times1\farcs3$. Following the relations in O'Dea \& Owen (1987), we obtain a total radio luminosity of $L_{rad}=1.9\times10^{41}$\,erg\,s$^{-1}$, a minimum magnetic field of $B_{min}\sim0.1$\,mG, minimum energy of $E_{min}=2.9\times10^{55}$\,ergs, and a minimum pressure of $P_{min}=1.2\times10^{-9}$\,dynes\,cm$^{-2}$.

The electron lifetime due to synchrotron radiative losses can be estimated using the relation, $t_{syn}\simeq33.4~B_{10}^{-3/2}~\nu_{c}^{-1/2}$ Myr, where $B_{10}$ is the magnetic field (``minimum'' $B$ field here) in units of 10\,$\mu$G, and $\nu_{c}$ is the critical frequency in GHz (e.g., Pacholczyk 1970). This turns out to be $\sim$3$\times10^5$ yrs for a critical frequency of 8\,GHz. Further assuming that the efficiency ($\epsilon$) with which the total jet energy is tapped to produce radio luminosity is 1\% (e.g., O'Dea 1985), the jet would have a kinetic luminosity ($L_{rad}$/$\epsilon$) of $\sim1.9\times10^{43}$\,erg\,s$^{-1}$. {Following the relation derived by Birzan et al. (2004) for radio-filled cavities, we also derive a jet kinetic luminosity of $\sim1.1\times10^{43}$\,erg\,s$^{-1}$.}

\section{Discussion}
\label{sec:discussion}
Our observations have revealed an apparent two-arm grand-design nuclear spiral
structure in \arp, extending from $\approx50$\,pc up to more than 500\,pc from the nucleus. We now discuss two possibilities for the origin of these spiral arms, namely that they are formed either through some dynamical process, or by jet-cloud interaction. 

\subsection{A Dynamical Formation Scenario}

The spiral structure that we find in our high resolution {\it HST}-ACS image could be described as a grand design nuclear spiral. Similar grand design nuclear spirals have been previously found around the nucleus of many Seyfert galaxies (Martini \& Pogge 1999). Maciejewski (2002, 2004b) argues that such grand-design spirals are shocks in the gas, leading to dissipation and inflow.

A strong correlation between the presence of an AGN and the presence of dusty nuclear spirals was found by Sim\~oes Lopes et al. (2007) in {\it HST} optical images
of a sample of early-type hosts. Pogge \& Martini (2002) also find nuclear spirals in almost all Seyfert galaxies of their sample. They also argue that there is  {\it ``a clear physical connection between the nuclear dust spirals on hundreds of parsec scales and large-scale bars and spiral arms in the host galaxies proper"}. In disk galaxies, theoretical works show how the presence of a non-axisymmetric potential is efficient in disturbing the circular orbits  of gas and stars, and to cause a loss of angular momentum due to torques and shocks (e.g., Schwarz 1984). This process  results in infall of matter from the outer parts towards the center of the gravitational potential (e.g., Combes \& Gerin 1985;  Knapen et al. 1995; Jogee et al. 2002). It is thus possible that large scale non-axisymmetric potentials lead to the formation of an inner structure, which in turn can lead the gas further inwards (e.g., Maciejewski \& Sparke 2002; Emsellem et al. 2001; Heller, Shlosman \& Englmaier 2001; Shlosman 2001; Maciejewski 2004a,b; Shlosman 2005). In the case of elliptical galaxies, there is indication that -- in at least half of them  -- the gas observed in the nuclear region comes from outside of the system (Sarzi et al. 2006). Thus, in elliptical galaxies with no sign of a gaseous disc, gas gets to the centres usually because a ballistic cloud undergoes dynamical friction and spirals in, and then cloud's angular momentum may spread it into a disc in the central kpc.

Nuclear spirals and filaments have been detected  via the contrast produced by the associated dust in broad-band images of essentially all nearby active galaxies (e.g. Malkan, Gorjian \& Tam 1998; Sim\~oes Lopes et al. 2007). For a few of them, mass inflows in ionized and molecular gas associated to nuclear spirals have been recently mapped down to a few parsecs or tens of parsecs of the AGN. Examples are the gas kinematic  studies around the active nuclei of the galaxies NGC\,7469 (Davies et al. 2004), NGC\,1097 (Fathi et al. 2005, 2006; Davies et al. 2009, van de Ven \& Fathi 2010), NGC\,6951 (Storchi-Bergmann et al. 2007), NGC\,4051 (Riffel et al. 08), NGC\,7582 (Riffel et al. 2009), NGC\,4151 (Storchi-Bergmann et al. 2010), NGC\,1068 (Mueller Sanchez et al. 2009) and M\,81 (Schnorr M\"uller et al 2011).

Similarly, the nuclear spiral in \arp\ may also trace a gas inflow. A possibility is that the current nuclear activity in \arp\ is being fueled by gas acquired from the companion  galaxy Arp\,102A, as \arp\ lies at the end of the tidal arms of Arp\,102A. This interaction may have fed gas into the galaxy and this gas may then have found its way to nuclear scales -- via the mechanisms discussed above -- to feed the AGN. We can calculate the amount of gas available for accretion using the H$\alpha$ flux from Halpern et al. (1996) as discussed in Section 3, of 4.3$\times 10^{-14}$\,erg\,cm$^{-2}$\,s$^{-1}$ and an adopted distance to \arp\ of 102\,Mpc, to obtain the H$\alpha$ luminosity of $L_{H\alpha}=5.4\times 10^{40}$erg\,s$^{-1}$. The ionized gas mass $M$ can be obtained (Osterbrock \& Ferland 2006) from
\begin{equation}
M=m_p\frac{L_{H\alpha}}{3.56\times 10^{-25}N_e}
\end{equation}
where $m_p$ is the mass of the proton in grams and $N_e$ is the electronic density in cm$^{-3}$. For a gas density of 100\,cm$^{-3}$ (estimated from the [SII]$\lambda6731/\lambda6717$ line ratio in Halpern et al. 1996), this gives a total ionized gas mass for the spiral of $M=1.3\times10^6$ solar masses. Considering the uncertainties in the H$\alpha$ flux as well as in the gas density, the resulting uncertainty in the calculated gas mass is a factor of $\approx$\,2.

\subsection{A Jet-Cloud Interaction}
The highly curved radio jet and its good overall spatial 
coincidence with the emission-line gas forming the eastern arm of the spiral suggests a scenario in which an initially relativistic, possibly precessing, radio jet shocks and
ionizes the surrounding gas. This is also suggested by the orientation of the
milliarcsecond VLBI structure, at PA$=40\degr$, which coincides with the orientation of the
inner part of the east spiral arm. In this case, the spiral arms could be formed 
from shocks which arise when the radio jet makes its way through the interstellar medium
(e.g., Veilleux, Tully, \& Bland-Hawthorn 1993; Veilleux, Cecil \& Bland-Hawthorn 2005). 
Both jet-like emission line structures and kinematically disturbed ionized gas  
associated with the radio jets have been found previously in a number of nearby 
Seyfert galaxies e.g., Mrk~6 (Capetti et al., 1995) and NGC\,1068 (Gallimore, Baum \& O'Dea 1996; Capetti et al. 1997; Axon et al. 1998). 

{It could be argued that the apparent bending of the jet in this ``radio quiet'' AGN, 
from a PA$\approx 40\degr$ on miliarcsecond scales to  $\approx 110\degr$ on arcsecond scales (Fig.~\ref{fig:HSTradio})  could be due to the ``deflection" of the radio jet by a ISM cloud, as observed  in other AGN such as NGC\,1068 (Gallimore, Baum \& O'Dea 1996). This is supported in the case of NGC\,1068 by a sharp change in the jet direction of  $\approx 20\degr$. However, in the \arp\ jet there is no sharp bend, and the ``smooth" continuous change of direction from pc to kpc-scales} can rather be attributed either to a viewing angle effect or precession. In the former case, the jet would have to be relativistic and at a relatively low ($< 45\degr$) inclination, which -- as we have already argued $-$ is consistent with a relativistic boosting explanation for its one-sidedness. Alternatively, the change in direction could result from jet precession, an effect that has been invoked to explain similar features in several Seyfert galaxies, for e.g., NGC\,3079 (Baan \& Irwin 1995), Mrk~6 (Kharb et al. 2006), NGC\,6764 (Kharb et al. 2010a), {and radio-loud AGN (Conway \& Murphy 1993; Kharb et al., 2010b)}, and may be a consequence either of instabilities in the accretion disk or the presence of a binary blackhole.

Finally, it could be also conjectured that the apparent alignment between the arcsec-scale-radio jet and the eastern arm is due to a chance projection of a foreshortened jet. This is very unlikely, because there is a very good correspondence between the east arm in the radio image and the one in the H$\alpha$ image. But our current data does not allow a more detailed discussion. 

\subsection{Ionization mechanism}
In section \ref{radio}, we estimated a kinetic luminosity for the radio jet of ($L_{rad}$/$\epsilon$) of $\sim1.9\times10^{43}$~erg~s$^{-1}$, which we can compare to the \Ha\ luminosity of the extended emission of 5.4$\times$\,10$^{40}$\, erg\,s$^{-1}$. This is much smaller than the kinetic luminosity of the jet, which is therefore energetically capable of ionizing the emission-line gas. The jet kinematic energy would be transferred to the gas via shocks. In a fast shock, (Sutherland, Bicknell \& Dopita 1993; Allen et al. 2008) optical emission lines are expected to arise both in the collisionally ionized gas forming the downstream cooling flow, and via photoionization of upstream gas by X-rays emitted by hot  gas immediately behind the shock. In either case, shock velocities $\ge 100$\,km\,$s^{-1}$ are required, implying comparable line widths for the emission arising in the shocked gas, while the widths of lines produced by the photoionized precursor would reflect the local velocity field of the undisturbed gas. For comparison, the narrow line FWHM, as measured by Halpern et al. (1996), are $\sim 400-600$\,km\,s$^{-1}$.

Alternatively, the emission line gas in the arms could be photoionized by the UV-X-ray continuum of the AGN itself. The optical-UV spectrum of Arp\,102B falls steeply into the UV ($f_\nu \propto \nu^{-2.1}$; Halpern et al. 1996) and thus contributes little flux above the Lyman limit. On the other hand, the $0.5-10$\,keV X-ray spectrum is well described by an absorbed power law of photon index $\Gamma\approx 1.6$ (Eracleous,  Halpern \& Charlton 2003), as is typical of Seyfert 1 galaxies. If this extends to the Lyman limit, it would provide sufficient ionizing photons to account for the \Ha\ emission in the arms. Assuming all photons with energies $13.6 \le h\nu \le 54.4$\,eV are absorbed by Hydrogen, the total \Ha\  flux is given by:

\begin{equation}
F_{H\alpha}= \frac{\alpha_{H\alpha}^{eff}}{\alpha_B} \left( \frac{\nu_{H\alpha}}{\nu_{1\,keV}}\right) F_{1\,keV}\int_{\nu_{HI}}^{\nu_{HeII}}\left( \frac{\nu}{\nu_{1\,keV}}\right)^{-(\alpha+1)}d\nu
\end{equation}

where $\alpha_{H\alpha}^{eff}$ and ${\alpha_B}$ are, respectively, the effective recombination coefficient for H$\alpha$ and the total case B recombination coefficient (Osterbrock \& Ferland 200)and the various frequencies are identified by their subscripts. Taking the 1\,keV flux given by Eracleous et al., $F_{1\,keV}\approx1.7~\mu$Jy, and $\alpha = \Gamma -1$, we find $F_{H\alpha}\sim 4\times10^{-14}$\,erg\,s$^{-1}$\,cm$^{-2}$, consistent with the observed \Ha\ flux. 

A detailed discussion of the line intensity ratios is beyond scope of the present paper. We merely note that LINER spectra have been interpreted both in terms of shocks (Dopita \& Sutherland 1995) and photoionization by a dilute, hard radiation field (e.g., Terashima, Ho, \& Ptak 2000; Ferland \& Netzer 1983). Stauffer et al. (1983) argued that shocks are most likely responsible for ionizing the narrow line region in \arp, citing among other things, the high electron temperature ($T>30,000\,K$) implied by the strength of the [O\,III]$\lambda 4363$ line, relative to [O\,III]$\lambda 5007$. However, this line is relatively weak and Halpern et al. note that their measured flux for this line is highly uncertain. Two-dimensional spectroscopy of the ``arms'' would help to distinguish between central source photoionization and shocks associated with the radio jet.

\section{Summary}
\label{sec:conclusions}
We report the discovery of a two-armed gaseous spiral structure in the central kiloparsec ($\approx$\,2\arcsec) of the E0 LINER/Seyfert 1 galaxy Arp\,102B in an {\it HST}-ACS \Ha\ emission-line gas image and adjacent continuum. 

Our morphological analysis has enabled us to trace the spiral arms from $0\farcs1$\,($\approx$\,50\,pc) to more than 1\arcsec\ ($\approx$\,500\,pc) from the nucleus to the east and west of the nucleus. Previous studies by our group have shown that spiral arms in the nuclear region of active galaxies can be channels to feed the SMBH, as in a few cases studied so far inflows have been found along such arms. In the case of \arp, a possible trigger for sending gas inwards could be its companion galaxy Arp\,102A. We calculate the mass of ionized gas of the spiral to be $M\approx10^{6}$ solar masses.
{We speculate that the recent accretion of gas to the AGN in Arp\,102B is related to its double-peaked emission-line profiles, presumably originating in the outer parts of a recently replenished  accretion disk.}

Nevertheless, archival VLA observations of Arp\,102B reveal  a one-sided radio jet structure which is well aligned with the east spiral arm, which suggests that the arms originate in a jet-cloud interaction. A milliarcsecond radio structure also aligns with the orientation of the inner part of the eastern spiral arm.
Thus, although the origin of the gas in the nuclear spiral may have been a recent accretion event, which probably triggered the nuclear activity producing the observed radio jet, the present observations suggest the alternative interpretation that the \Ha\ spiral arms are a result of the interaction between the jet and the remaining circum-nuclear gas.

The kinetic energy delivered by the jet is sufficient to ionize the \Ha\ emitting gas. However, it could also be photoionized by  AGN X-ray continuum; the morphological relationship could simply be due to an increased emission measure where shocks have compressed the surrounding gas.  The absence of a radio counter-jet coincident with the western arm seen in the \Ha\ image can be explained if the jet is relativistic. This is further supported  by the milliarcsecond  structure which shows only one extranuclear component at PA\,$\approx$\,40\,$\deg$. An estimate for the relativistic particle speed in the inner part of the jet is 0.45$c$. 

\section*{Acknowledgments}
Based on observations with the NASA/ESA Hubble Space Telescope obtained at 
the Space Telescope Science Institute, which is operated by the Association 
of Universities for Research in Astronomy, Inc., under NASA contract NAS5-26555. 
The National Radio Astronomy Observatory is a facility of the National Science Foundation operated under cooperative agreement by Associated Universities, Inc.
We thank an anonymous referee for a careful reading of the manuscript and useful suggestions which helped to improve the paper.
TSB acknowledges support from the Brazilian institution CNPq. 
KF acknowledges financial support from NASA through grant number 
{\it HST}-GO 09782.01, from the Royal Swedish Academy of Sciences and 
from the Swedish Research Council.

\end{document}